\documentclass[sigconf,nonacm]{acmart}

\setcopyright{acmcopyright}
\copyrightyear{2022}
\acmYear{2022}
\acmDOI{XXXXXXX.XXXXXXX}

\acmConference[FAccTRec '22]{5th FAccTRec Workshop on Responsible Recommendation}{September 2022}{Seattle, WA}
\acmBooktitle{Woodstock '18: ACM Symposium on Neural Gaze Detection,
 June 03--05, 2018, Woodstock, NY} 
\usepackage{listings}
\usepackage{graphicx} 
\usepackage{natbib} 
\usepackage{url}

\begin{document}

\title{Let's Learn from Children: Scaffolding to Enable Search as Learning in the Educational Environment}
\titlenote{Presented at IWILDS 2022.}
\fancyhead[LE,RO]{\small\sffamily IWILDS 2022}

\author{Monica Landoni}
\authornote{Authors contributed equally to this paper.}
\email{monica.landoni@usi.ch}
\affiliation{%
  \institution{Universit\`a della Svizzera Italiana}
  \city{Lugano}
  \country{Switzerland}
  }
\author{Maria Soledad Pera}
\authornotemark[2]
\orcid{0000-0002-2008-9204}
\email{solepera@boisestate.edu}
\affiliation{%
  \institution{Web Information Systems Group \\ Delft University of Technology}
  \city{Delft}
  \country{The Netherlands}
  }
\author{Emiliana Murgia}
\authornotemark[2]
\email{emiliana.murgia@unimib.it}
\affiliation{%
 \institution{Universit\`a degli Studi di Milano-Bicocca}
 \country{Milan, Italy}
 }
 \author{Theo Huibers}
 \authornotemark[2]
\email{t.w.c.huibers@utwente.nl}
\affiliation{%
  \institution{University of Twente}
  \city{Enschede}
  \country{The Netherlands}
  }  
  
\renewcommand{\shortauthors}{Landoni, Pera, Murgia, and Huibers}

\begin{abstract}
In this manuscript, we argue for the need to further look at search as learning (SAL) with children as the primary stakeholders. Inspired by how children learn and considering the classroom (regardless of the teaching modality) as a natural educational ecosystem, we posit that scaffolding is the tie that can simultaneously allow for learning to search while searching for learning. The main contribution of this work is a list of open challenges focused on the primary school classroom for the IR community to consider when setting up to explore and make progress on SAL research with and for children and beyond. 
\end{abstract}


\maketitle

\section{How Children Learn}
\label{sec:Scaffolding}
Searching for information on a broad range of topics using different devices and modalities (e.g., traditional text-based to voice-assisted) is habitual. For some users, particularly those who are part of the `Google generation' (born after 1993), it is evident that the path to gaining knowledge is through the use of the internet and search engines \cite{rowlands2008google}. 

We bring our attention to a specific subgroup of the Google generation: \textit{children}\footnote{As per UNICEF's definition, we consider children individuals up to the age of 18.}. They prefer using popular, commercial search engines to satisfy their information needs \cite{jochmann2010children,rowlands2008google}. Despite many attempts, these tools have not been calibrated to serve this population, resulting in well-documented barriers children encounter \cite{azpiazu2017online}. 
Nonetheless, we see reports in the literature of how resilient children can be in identifying ways to circumvent barriers and still use search engines to address their leisure-related search tasks. They do so, for instance, by adopting different search roles \cite{druin2010children}, each resulting in varying degrees of satisfaction when conducting search tasks individually in the home setting. Looking at the dynamics of information acquisition in the classroom, recent publications disclose how the natural scaffolding provided by teachers and peers is fundamental in supporting children in their quest for information necessary to solve curriculum-related inquiry tasks \cite{landoni2021getting,milton2019here}.  

In closing the gap between what children can do on their own and what they can do with help, and inspired by Vygotsky's theory of Zone of Proximal Development \cite{vygotsky1978mind} as well as the problem-based learning paradigm \cite{allen2011problem}, we regard \textit{scaffolding} as one of the most effective ways to support children in their development of skills and knowledge. This is both \textit{explicit} (training) in the form of search and media literacy instruction and teacher interventions, as well as \textit{implicit} (algorithms) via integrated added functionality (e.g., query formulation support) until children are ready to independently and effectively use search engines. While exploring how different forms of scaffolding can provide a better search experience for children, we believe it is also possible to recognise patterns to help improve search tools for adults. In other words, the design and development of better search solutions for adults could be guided by how children take advantage of scaffolding and so learn from their experience.

\section{Search as Learning}
In the late 1990s, Marchionini defined the information-seeking process as ``a process, in which humans purposefully engage to
change their state of knowledge” \cite{marchionini1997information}. Following this line of thinking, researchers in the early 2000s focused on the \textit{search as learning} (\textbf{SAL}) paradigm. Research in this area discusses both on \textit{learning to search} (skill development) or \textit{searching to learn} (interaction with information) \cite{hansen2016recent,rieh2016towards}.

Literature in the SAL area is broad. It covers different aspects: (i) the connection of search behaviour and learning outcomes \cite{ghosh2018searching}, (ii) domain knowledge and its impact on learning outcomes when searching \cite{o2020role}, (iii) how to measure learning within SAL environments \cite{urgo2022learning}, (iv) datasets that sustain further research in this area \cite{otto2022sal}, (v) new frameworks to better understanding and measuring of SAL processes \cite{liu2021interest}, (vi) exploring the use of active reading tools to support search as learning \cite{roy2021note}, and (vii) knowledge gain as a result of learning-related inquiry tasks \cite{liu2019investigation}, to name a few. (An in-depth review of SAL-related literature can be found in \cite{hoppe2018current,kameni2022search}).
Early works related to SAL, including the Dagstuhl manifesto \cite{sal-dagstuhl} and the first SAL workshop co-located with ACM SIGIR \cite{sal-workshop16}, mention non-typical user groups, e.g., low-literacy searchers, non-experts, and children. Yet, most research works target or refer to studies involving more mature users. 

Children are a user group whose cognitive and literacy skills are in the development stages. Moreover, they are first exposed to search engines without prior knowledge and thus must learn how to use them (beyond intuitive engagement that might not account for best practices). Consequently, it becomes imperative to understand their SAL needs and respond accordingly, rather than assuming that SAL outcomes observed from adults are directly applicable to children.
Vice versa, we posit that search solutions that prove effective with children could also aid adult searchers.  

\section{SAL and Children}
\label{sec:SAl-kids}
Among research efforts connecting SAL and children, we find the works by Downs et al. \cite{downs2020guiding,downs2021kidspell} who focus on how children express their information needs to prompt the search process. In particular, they introduce KidSpell, a spellchecker based on phonetic keys that explicitly accounts for the spelling mistakes children make. The learning to search perspective is accounted for by the visual cues that help children identify potential misspellings and how to correct them to compose more effective queries.

Madrazo Azpiazu et al. \cite{azpiazu2016finding, azpiazu2017online} bring attention to both of SAL's perspectives (i.e., searching for learning and learning to search), by discussing the architecture of a search environment aiming to support children conducting information discovery tasks in the classroom. Similarly, Landoni et al. \cite{landoni2022have,landoni2021somewhere,landoni2021right} also address both of SAL's perspectives. They explore the use of visual cues as a means to support learning to identify relevant resources--a key aspect of the search process--retrieved in response to children's queries. The authors also explore search behavior and grades children (ages 8 to 11) obtained in an inquiry assignment for which they used search engines to locate information related to `ancient Rome', a topic presented as part of class instruction.

Von Hoyer et al. \cite{von2022search} introduce a novel framework that leverages both computer science and psychology perspectives to model SAL. Even though this work does not explicitly focus on children, the authors use the interaction of an adolescent seeking information on the Internet to complete a homework assignment to illustrate the different components of their proposed framework. This serves as another indication of the importance of bringing attention to SAL and how it connects to children.

\section{SAL and Scaffolding: Open Challenges}
The needs and expectations of children when searching vary based on their cognitive abilities, prior exposure, topics of interest, and many other aspects. Further, Usta et al. \cite{usta2014k} emphasise that children's search behaviour in the classroom differs from that observed for general web searches. In an educational environment\footnote{In this work, we use `classroom' and `learning context' interchangeably to refer to the educational environment, regardless of the teaching modality, i.e., searching for information related to the school curriculum in a traditional classroom or online from home. 
}, searching is directly related to learning, as children search for information that enables them to complete curriculum-related assignments. In contrast, search engines children utilise are \textit{
``optimised for acquiring factual
knowledge but are less successful at facilitating other kinds
of learning, such as understanding, analysis, application or
synthesis, in terms of Bloom's taxonomy"} \cite{taibi2017sar}. Motivated by these considerations and to control the scope as we set a strong foundation for SAL research targeting children, we focus on the educational environment. 

As previously mentioned, scaffolding is extensively adopted in the classroom context. One of the most common scaffolding practices involves the use of mediators as physical or visual objects, in addition to actions, to help children to learn and recall something \cite{berk1995scaffolding}. With this in mind, we posit that when children are the primary stakeholders, \textit{scaffolding}--temporary adaptive support \cite{shvarts2019early}--could serve as a conduit to connect both of SAL's perspectives and focus on learning to search while searching for learning. Informed by our prior studies, along with associated findings pertaining to the needs of young searchers when conducting online information discovery tasks for the classroom, we identify a number of challenges that need  consideration: the use of search tools in a learning context \cite{landoni2021right}, the need for evaluation for multiple perspectives \cite{landoni2021somewhere}, the individual needs of young searchers \cite{landoni2021getting}, the need for voice-driven search \cite{landoni2019sonny}, ethical considerations \cite{landoni2022ethical}, and the possible impact of the COVID-19 pandemic \cite{clef22}.
We outline open directions that can advance understanding of the use of scaffolding to simultaneously support enhanced search experiences when utilising search tools while enabling search literacy\footnote{We consider best practices in the use of search tools and associated theory, such as concerns related to misinformation, as per the definition of web search literacy which states that ``Web search literacy as the ability to identify, locate and
effectively use information on the Web" \cite{karatassis2017websail}.} advancement in an educational environment.

\subsection{How does SAL looks and feels in the classroom context?}
As previously stated, children favour commercial search engines. Hence, algorithmic functionality providing the scaffolding is needed so that children can successfully traverse through each of the stages of the information-seeking process become a must. Informed by findings reported in \cite{landoni2021right}, open problems in this area include, for example, re-ranking algorithms that can prioritise retrieved results to match the classroom context \cite{milton2020ranking} and simplifying the text complexity of resources so that users can read and understand based on their reading skills. Query formulation support and query suggestion strategies tailored to the classroom context are also of interest.

The intent, however, is not for children to just use search engines but instead for them to acquire search literacy knowledge along the way. Hence, scaffolding should also incorporate  novel interfaces that can offer clues on the relevance of retrieved resources, guidance on navigation practices, best practices in the query suggestion section, and explanation of results that have the potential to be misinformation. This latter is particularly challenging for children to discern \cite{rouet2018reading}. 

Note that open problems in this area could complement those mentioned by von Hoyer et al. \cite{von2022search} in describing their proposed framework (see Section \ref{sec:SAl-kids}).

\subsection{How do we measure performance?}

Evaluation remains of at most importance. Assessing the performance of search systems serving children is challenging, given the lack of benchmarks, datasets to compare and contrast performance, metrics to use to measure performance, and unambiguous definitions of what it means for a retrieved result or systems to be `good' for children \cite{huibers2021does, milton2020evaluating,anuyah2019need}. 

We argue that in the classroom context, evaluation is even more challenging. We start from getting a better understanding of how children interpret relevance in the classroom context as described in \cite{landoni2021somewhere,landoni2022ethical}. There is also a need for assessing how available interfaces support children in performing their search tasks as discussed in \cite{aliannejadi2021children} where we introduce four lenses grounded on UX factors. In the SAL context, (simultaneous) assessment of `learning' both in terms of search literacy and knowledge gain as a result of searching remains an open problem. This requires the involvement of interdisciplinary researchers--beyond computer science--that can outline new (performance) metrics as well as design short and long-term studies involving children in their natural classroom environment as opposed to a lab setting.

\subsection{Why look at children as individuals, rather than `just' as a user group?}
We have recently explored the different roles children play when searching in the classroom  \cite{landoni2021getting}. Our findings revealed that the complexity of the search task, its topic, and other reading and comprehension abilities, to name a few, could condition children's behaviour when searching for information related to the classroom curriculum. This prompts the need for scaffolding to react to each of these roles instead of adopting a one-size-fits-all approach.  In a similar line of investigation, adequately accounting for children's unique abilities (in development) is also challenging. For example, a child affected by dyslexia would need different scaffolding for query formulation, result exploration, and distilling information from retrieved resources to support learning. Scaffolding addressing such a concern would differ from the one aiding children for whom processing text is less of a challenge. In the context of SAL, this is a far-reaching issue, as it applies to adult users as well.  

\subsection{What about interactions beyond text?}
The modality in which children access search systems is worth exploring. We have discussed some of the limitations children experience when engaging with traditional text-based search engines. Yet, it is worth mentioning that younger searchers find it easier to turn to voice-based devices like Siri or Alexa \cite{lovato2015siri}. In this case, what would scaffolding look like? How to support learning as a result of searching when most often voice-based devices would directly offer an answer to users' prompts? How does scaffold learning to formulate queries that Siri or Alexa can understand?  

Moving beyond typical searches for websites,and based on outcomes observed as a result of the study reported in \cite{landoni2019sonny} pertaining how children engage with search companions, what are the open problems about SAL when children seek videos (e.g., turn to YouTube \cite{neumann2020young}) that can support their learning? In this case, what would be the scaffolding needed to discern educational and suitable retrieved videos from those that do not align with the classroom context? Should the support offered for query formulation, resource prioritisation, and relevance assessment differ from that devised for more traditional searches? Or should instead  scaffolding resemble strategies teachers adopt when searching for videos for their students \cite{fyfield2021navigating}?

\subsection{How to reconcile research advancement and the rights of the child?} 
Design, development, assessment, and deployment of search systems that can offer the implicit and explicit scaffolding needed to enable SAL for children in the classroom to depend on the availability of data that facilitates modelling of users and their interactions. Further, once deployed, search systems with scaffolding embedded must adapt over time, reacting to individual users' needs. Research advancement and personalization have direct implications on children and their rights, such as privacy, i.e., the protection of their data \cite{thinkautomation_2020}. (For an overview on this matter see \cite{JRC127564}). 

Already researchers, practitioners, and other governmental organisations bring attention to the degree to which Artificial Intelligence (AI) impacts children, particularly in the classroom context, along with open challenges in this area \cite{henriques2020children, unicef-ai-kids}. AI aims to build machines and software that can mimic intelligent human behaviour \cite{unicef-ai-kids}. Systems that deliver the level of scaffolding that teachers would naturally provide their students, i.e., systems that automatically adapt and respond to the learning needs of children that change over time, are at the core of AI research. This is why SAL researchers and practitioners should be mindful of ethical implications \cite{landoni2022ethical} and the multiple aspects (not just technology) impacting design for children as they address some of the open problems discussed in the aforementioned sections of this manuscript.

\subsection{Is there a connection between the COVID-19 pandemic and SAL research?}

As we discuss in \cite{clef22}, over the last three years, the classroom context has been severely affected by the COVID-19 pandemic. This impact has shifted across the years, from complete lock-downs forcing learning to happen online to hybrid interactions. Studies conducted to understand the implications of the pandemic on children and learning are ongoing, but some shreds of evidence emerge from early on \cite{lee2020mental,singh2020impact}.
Of note is how students have lowered their autonomy level in completing a task, regulating their attention, and independent learning. The pandemic forced them to learn from home where they had the constant presence of adults ready to scaffold their learning experience even when they did not need to.

The task for teachers now is to choose the best strategies and methodologies to bring students back to the level of independence they exhibited pre-pandemic. For SAL researchers, the challenge is specifically on avoiding the perpetuation of this lack of independence as much as on discerning needed aid (e.g. embedded instruction and algorithmic functionality). Doing so will undoubtedly require new studies to observe how children engage with search tasks pertaining to the educational environment and, more importantly, involving teachers. Teachers as partners for SAL-related research will enable a better interpretation of which type of scaffolding is needed post-pandemic. Moreover, the teachers will share their perceptions of how students would fare when presented with scaffolding designed to aid their pre-pandemic skill set. Will students take advantage of it or would they simply bypass it in favour of more guidance afforded directly by teachers or peers?

\section{Concluding Remarks}
With this work, we aimed to bring attention to the need to advance knowledge about SAL for children. As a step towards achieving that goal, we scoped our discussion to the educational environment, as the perfect ecosystem where searching for learning and learning to search coexist. We outlined directions that call for the participation of researchers and industry practitioners with varied expertise (e.g., Information Retrieval, Information Science, Human-Computer Interaction, Education, Natural-Language Understanding, and Child Development)  so that resulting outcomes have theoretical, but more importantly, practical implications in real-world scenarios.

If learning to learn is an essential skill in our society, then SAL is vital to sustaining children and adults in their continued learning process. As such, it needs more attention and study. Here, we are not providing answers, only open challenges and possible directions for future research to guide us in the quest for better SAL experiences for children. The connection between scaffolding and SAL has been mentioned in prior literature with adults as primary stakeholders. Most notably, Camara et al \cite{camara2021searching} report that scaffolding in the form of curated topic lists or exploration feedback did not influence learning outcomes, it did impact their search behaviour. This evidences the value of expanding research explorations on the natural connection between SAL and scaffolding. In our case, we strongly believe that a deep exploration of the role scaffolding plays in the classroom could support its adoption as an essential functionality of search tools to provide a better experience \textit{for all}.

\bibliographystyle{IEEEtran} 
\bibliography{References}

\begin{thebibliography}{10}
\providecommand{\url}[1]{#1}
\csname url@samestyle\endcsname
\providecommand{\newblock}{\relax}
\providecommand{\bibinfo}[2]{#2}
\providecommand{\BIBentrySTDinterwordspacing}{\spaceskip=0pt\relax}
\providecommand{\BIBentryALTinterwordstretchfactor}{4}
\providecommand{\BIBentryALTinterwordspacing}{\spaceskip=\fontdimen2\font plus
\BIBentryALTinterwordstretchfactor\fontdimen3\font minus
  \fontdimen4\font\relax}
\providecommand{\BIBforeignlanguage}[2]{{%
\expandafter\ifx\csname l@#1\endcsname\relax
\typeout{** WARNING: IEEEtran.bst: No hyphenation pattern has been}%
\typeout{** loaded for the language `#1'. Using the pattern for}%
\typeout{** the default language instead.}%
\else
\language=\csname l@#1\endcsname
\fi
#2}}
\providecommand{\BIBdecl}{\relax}
\BIBdecl

\bibitem{rowlands2008google}
I.~Rowlands, D.~Nicholas, P.~Williams, P.~Huntington, M.~Fieldhouse, B.~Gunter,
  R.~Withey, H.~R. Jamali, T.~Dobrowolski, and C.~Tenopir, ``The google
  generation: the information behaviour of the researcher of the future,'' in
  \emph{Aslib proceedings}.\hskip 1em plus 0.5em minus 0.4em\relax Emerald
  Group Publishing Limited, 2008.

\bibitem{jochmann2010children}
H.~Jochmann-Mannak, T.~Huibers, L.~Lentz, and T.~Sanders, ``Children searching
  information on the internet: Performance on children’s interfaces compared
  to google,'' in \emph{SIGIR}, vol.~10, 2010, pp. 27--35.

\bibitem{azpiazu2017online}
I.~M. Azpiazu, N.~Dragovic, M.~S. Pera, and J.~A. Fails, ``Online searching and
  learning: Yum and other search tools for children and teachers,''
  \emph{Information Retrieval Journal}, vol.~20, no.~5, pp. 524--545, 2017.

\bibitem{druin2010children}
A.~Druin, E.~Foss, H.~Hutchinson, E.~Golub, and L.~Hatley, ``Children's roles
  using keyword search interfaces at home,'' in \emph{Proceedings of the SIGCHI
  Conference on Human Factors in Computing Systems}, 2010, pp. 413--422.

\bibitem{landoni2021getting}
M.~Landoni, T.~Huibers, M.~Aliannejadi, E.~Murgia, and M.~S. Pera, ``Getting to
  know you: Search logs and expert grading to define children’s search roles
  in the classroom,'' in \emph{2nd International Conference on Design of
  Experimental Search and Information REtrieval Systems, DESIRES 2021}, 2021,
  pp. 44--52.

\bibitem{milton2019here}
A.~Milton, E.~Murgia, M.~Landoni, T.~Huibers, and M.~S. Pera, ``Here, there,
  and everywhere: Building a scaffolding for children’s learning through
  recommendations,'' in \emph{Proceedings of the 1st Workshop on the Impact of
  Recommender Systems co-located with the 13th ACM Conference on Recommender
  Systems (ACM RecSys 2019)}, 2019, available at:
  \url{http://ceur-ws.org/Vol-2462/short2.pdf}.

\bibitem{vygotsky1978mind}
L.~S. Vygotsky and M.~Cole, \emph{Mind in society: Development of higher
  psychological processes}.\hskip 1em plus 0.5em minus 0.4em\relax Harvard
  university press, 1978.

\bibitem{allen2011problem}
D.~E. Allen, R.~S. Donham, and S.~A. Bernhardt, ``Problem-based learning,''
  \emph{New directions for teaching and learning}, vol. 2011, no. 128, pp.
  21--29, 2011.

\bibitem{marchionini1997information}
G.~Marchionini, \emph{Information seeking in electronic environments}.\hskip
  1em plus 0.5em minus 0.4em\relax Cambridge university press, 1997, no.~9.

\bibitem{hansen2016recent}
P.~Hansen and S.~Y. Rieh, ``Recent advances on searching as learning: An
  introduction to the special issue,'' \emph{Journal of Information Science},
  vol.~42, no.~1, pp. 3--6, 2016.

\bibitem{rieh2016towards}
S.~Y. Rieh, K.~Collins-Thompson, P.~Hansen, and H.-J. Lee, ``Towards searching
  as a learning process: A review of current perspectives and future
  directions,'' \emph{Journal of Information Science}, vol.~42, no.~1, pp.
  19--34, 2016.

\bibitem{ghosh2018searching}
S.~Ghosh, M.~Rath, and C.~Shah, ``Searching as learning: Exploring search
  behavior and learning outcomes in learning-related tasks,'' in
  \emph{Proceedings of the 2018 conference on human information interaction \&
  retrieval}, 2018, pp. 22--31.

\bibitem{o2020role}
H.~L. O'Brien, A.~Kampen, A.~W. Cole, and K.~Brennan, ``The role of domain
  knowledge in search as learning,'' in \emph{Proceedings of the 2020
  Conference on Human Information Interaction and Retrieval}, 2020, pp.
  313--317.

\bibitem{urgo2022learning}
K.~Urgo and J.~Arguello, ``Learning assessments in search-as-learning: A survey
  of prior work and opportunities for future research,'' \emph{Information
  Processing \& Management}, vol.~59, no.~2, p. 102821, 2022.

\bibitem{otto2022sal}
C.~Otto, M.~Rokicki, G.~Pardi, W.~Gritz, D.~Hienert, R.~Yu, J.~von Hoyer,
  A.~Hoppe, S.~Dietze, P.~Holtz \emph{et~al.}, ``Sal-lightning dataset: Search
  and eye gaze behavior, resource interactions and knowledge gain during web
  search,'' \emph{arXiv preprint arXiv:2201.02339}, 2022.

\bibitem{liu2021interest}
J.~Liu and Y.~J. Jung, ``Interest development, knowledge learning, and
  interactive ir: Toward a state-based approach to search as learning,'' in
  \emph{Proceedings of the 2021 Conference on Human Information Interaction and
  Retrieval}, 2021, pp. 239--248.

\bibitem{roy2021note}
N.~Roy, M.~V. Torre, U.~Gadiraju, D.~Maxwell, and C.~Hauff, ``Note the
  highlight: Incorporating active reading tools in a search as learning
  environment,'' in \emph{Proceedings of the 2021 Conference on Human
  Information Interaction and Retrieval}, 2021, pp. 229--238.

\bibitem{liu2019investigation}
H.~Liu, C.~Liu, and N.~J. Belkin, ``Investigation of users' knowledge change
  process in learning-related search tasks,'' \emph{Proceedings of the
  Association for Information Science and Technology}, vol.~56, no.~1, pp.
  166--175, 2019.

\bibitem{hoppe2018current}
A.~Hoppe, P.~Holtz, Y.~Kammerer, R.~Yu, S.~Dietze, and R.~Ewerth, ``Current
  challenges for studying search as learning processes,'' in \emph{7th Workshop
  on Learning \& Education with Web Data (LILE2018), in conjunction with ACM
  Web Science}, 2018.

\bibitem{kameni2022search}
J.~S. Kameni, B.~Batchakui, and R.~Nkambou, ``Search engines in learning
  contexts: A literature review.'' \emph{International Journal of Emerging
  Technologies in Learning}, vol.~17, no.~2, 2022.

\bibitem{sal-dagstuhl}
K.~Collins-Thompson, P.~Hansen, and C.~Hauff, ``Search as learning (dagstuhl
  seminar 17092),'' in \emph{Dagstuhl reports}, vol.~7, no.~2.\hskip 1em plus
  0.5em minus 0.4em\relax Schloss Dagstuhl-Leibniz-Zentrum fuer Informatik,
  2017.

\bibitem{sal-workshop16}
J.~Gwizdka, P.~Hansen, C.~Hauff, J.~He, and N.~Kando, ``Search as learning
  (sal) workshop 2016,'' in \emph{Proceedings of the 39th International ACM
  SIGIR conference on Research and Development in Information Retrieval}, 2016,
  pp. 1249--1250.

\bibitem{downs2020guiding}
B.~Downs, A.~Shukla, M.~Krentz, M.~S. Pera, K.~L. Wright, C.~Kennington, and
  J.~Fails, ``Guiding the selection of child spellchecker suggestions using
  audio and visual cues,'' in \emph{Proceedings of the Interaction Design and
  Children Conference}, 2020, pp. 398--408.

\bibitem{downs2021kidspell}
B.~Downs, M.~S. Pera, K.~L. Wright, C.~Kennington, and J.~A. Fails, ``Kidspell:
  Making a difference in spellchecking for children,'' \emph{International
  Journal of Child-Computer Interaction}, p. 100373, 2021.

\bibitem{azpiazu2016finding}
I.~M. Azpiazu, N.~Dragovic, and M.~S. Pera, ``Finding, understanding and
  learning: Making information discovery tasks useful for children and
  teachers,'' in \emph{SAL@ SIGIR}, 2016.

\bibitem{landoni2022have}
M.~Landoni, M.~Aliannejadi, T.~Huibers, E.~Murgia, and M.~S. Pera, ``Have a
  clue! the effect of visual cues on children’s search behavior in the
  classroom,'' in \emph{ACM SIGIR Conference on Human Information Interaction
  and Retrieval}, 2022, pp. 310--314.

\bibitem{landoni2021somewhere}
M.~Landoni, T.~Huibers, E.~Murgia, M.~Aliannejadi, and M.~S. Pera, ``Somewhere
  over the rainbow: Exploring the sense for relevance in children,'' in
  \emph{European Conference on Cognitive Ergonomics 2021}, 2021, pp. 1--5.

\bibitem{landoni2021right}
M.~Landoni, M.~Aliannejadi, T.~Huibers, E.~Murgia, and M.~S. Pera, ``Right way,
  right time: Towards a better comprehension of young students’ needs when
  looking for relevant search results,'' in \emph{Proceedings of the 29th ACM
  Conference on User Modeling, Adaptation and Personalization}, 2021, pp.
  256--261.

\bibitem{von2022search}
J.~Von~Hoyer, A.~Hoppe, Y.~Kammerer, C.~Otto, G.~Pardi, M.~Rokicki, R.~Yu,
  S.~Dietze, R.~Ewerth, and P.~Holtz, ``The search as learning spaceship:
  Toward a comprehensive model of psychological and technological facets of
  search as learning.'' \emph{Frontiers in Psychology}, vol.~13, pp.
  827\,748--827\,748, 2022.

\bibitem{usta2014k}
A.~Usta, I.~S. Altingovde, I.~B. Vidinli, R.~Ozcan, and {\"O}.~Ulusoy, ``How
  k-12 students search for learning? analysis of an educational search engine
  log,'' in \emph{Proceedings of the 37th international ACM SIGIR conference on
  Research \& development in information retrieval}, 2014, pp. 1151--1154.

\bibitem{taibi2017sar}
D.~Taibi, G.~Fulantelli, I.~Marenzi, W.~Nejdl, R.~Rogers, and A.~Ijaz,
  ``Sar-web: a semantic web tool to support search as learning practices and
  cross-language results on the web,'' in \emph{2017 IEEE 17th International
  Conference on Advanced Learning Technologies (ICALT)}.\hskip 1em plus 0.5em
  minus 0.4em\relax IEEE, 2017, pp. 522--524.

\bibitem{berk1995scaffolding}
L.~E. Berk and A.~Winsler, \emph{Scaffolding Children's Learning: Vygotsky and
  Early Childhood Education. NAEYC Research into Practice Series. Volume
  7.}\hskip 1em plus 0.5em minus 0.4em\relax ERIC, 1995.

\bibitem{shvarts2019early}
A.~Shvarts and A.~Bakker, ``The early history of the scaffolding metaphor:
  Bernstein, luria, vygotsky, and before,'' \emph{Mind, Culture, and Activity},
  vol.~26, no.~1, pp. 4--23, 2019.

\bibitem{landoni2019sonny}
M.~Landoni, D.~Matteri, E.~Murgia, T.~Huibers, and M.~S. Pera, ``Sonny, cerca!
  evaluating the impact of using a vocal assistant to search at school,'' in
  \emph{International conference of the cross-language evaluation forum for
  European languages}.\hskip 1em plus 0.5em minus 0.4em\relax Springer, 2019,
  pp. 101--113.

\bibitem{landoni2022ethical}
M.~Landoni, T.~Huibers, E.~Murgia, and M.~S. Pera, ``Ethical implications for
  children’s use of search tools in an educational setting,''
  \emph{International Journal of Child-Computer Interaction}, vol.~32, p.
  100386, 2022.

\bibitem{clef22}
M.~S. Pera, M.~Aliannejadi, M.~Landoni, E.~Murgia, and T.~Huibers, ``How do
  children imagine a search companion for the classroom?'' in
  \emph{International conference of the cross-language evaluation forum for
  European languages}.\hskip 1em plus 0.5em minus 0.4em\relax Springer, 2022,
  p. to appear.

\bibitem{karatassis2017websail}
I.~Karatassis, ``Websail: Computer-based methods for enhancing web search
  literacy,'' in \emph{Proceedings of the 2017 conference on conference human
  information interaction and retrieval}, 2017, pp. 403--405.

\bibitem{milton2020ranking}
A.~Milton, O.~Anuya, L.~Spear, K.~L. Wright, and M.~S. Pera, ``A ranking
  strategy to promote resources supporting the classroom environment,'' in
  \emph{2020 IEEE/WIC/ACM International Joint Conference on Web Intelligence
  and Intelligent Agent Technology (WI-IAT)}.\hskip 1em plus 0.5em minus
  0.4em\relax IEEE, 2020, pp. 121--128.

\bibitem{rouet2018reading}
J.-F. Rouet and A.~Potocki, ``From reading comprehension to document literacy:
  learning to search for, evaluate and integrate information across texts/de la
  lectura a la alfabetizaci{\'o}n documental: aprender a buscar, evaluar e
  integrar informaci{\'o}n de diversos textos,'' \emph{Infancia y Aprendizaje},
  vol.~41, no.~3, pp. 415--446, 2018.

\bibitem{huibers2021does}
T.~Huibers, M.~Landoni, M.~S. Pera, J.~A. Fails, E.~Murgia, and N.~Kucirkova,
  ``What does good look like? report on the 3 rd international and
  interdisciplinary perspectives on children \& recommender and information
  retrieval systems (kidrec) at idc 2019,'' in \emph{ACM SIGIR Forum}, vol.~53,
  no.~2.\hskip 1em plus 0.5em minus 0.4em\relax ACM New York, NY, USA, 2021,
  pp. 76--81.

\bibitem{milton2020evaluating}
A.~Milton and M.~S. Pera, ``Evaluating information retrieval systems for
  kids,'' \emph{4th International and Interdisciplinary Perspectives on
  Children \& Recommender and Information Retrieval Systems (KidRec '20),
  co-located with the 19th ACM International Conference on Interaction Design
  and Children (IDC '20). arXiv preprint arXiv:2005.12992}, 2020.

\bibitem{anuyah2019need}
O.~Anuyah, M.~Green, A.~Milton, and M.~Pera, ``The need for a comprehensive
  strategy to evaluate search engine performance in the classroom,'' in
  \emph{3rd KidRec Workshop co-located with ACM IDC}, 2019.

\bibitem{aliannejadi2021children}
M.~Aliannejadi, M.~Landoni, T.~Huibers, E.~Murgia, and M.~S. Pera, ``Children's
  perspective on how emojis help them to recognise relevant results: Do actions
  speak louder than words?'' in \emph{Proceedings of the 2021 Conference on
  Human Information Interaction and Retrieval}, 2021, pp. 301--305.

\bibitem{lovato2015siri}
S.~Lovato and A.~M. Piper, ``" siri, is this you?" understanding young
  children's interactions with voice input systems,'' in \emph{Proceedings of
  the 14th international conference on interaction design and children}, 2015,
  pp. 335--338.

\bibitem{neumann2020young}
M.~M. Neumann and C.~Herodotou, ``Young children and youtube: A global
  phenomenon,'' \emph{Childhood Education}, vol.~96, no.~4, pp. 72--77, 2020.

\bibitem{fyfield2021navigating}
M.~Fyfield, M.~Henderson, and M.~Phillips, ``Navigating four billion videos:
  teacher search strategies and the youtube algorithm,'' \emph{Learning, Media
  and Technology}, vol.~46, no.~1, pp. 47--59, 2021.

\bibitem{thinkautomation_2020}
\BIBentryALTinterwordspacing
``Ai for children: The risks and the rights,'' Oct 2020. [Online]. Available:
  \url{https://www.thinkautomation.com/automation-ethics/ai-for-children-the-risks-and-the-rights/.}
\BIBentrySTDinterwordspacing

\bibitem{JRC127564}
\BIBentryALTinterwordspacing
V.~Charisi, S.~Chaudron, R.~D. Gioia, R.~Vuorikari, M.~E. Planas, J.~S. Martin,
  and E.~G. Gutierrez, ``Artificial intelligence and the rights of the child :
  Towards an integrated agenda for research and policy,'' Luxembourg
  (Luxembourg), Scientific analysis or review KJ-NA-31048-EN-N (online), 2022.
  [Online]. Available:
  \url{https://publications.jrc.ec.europa.eu/repository/handle/JRC127564}
\BIBentrySTDinterwordspacing

\bibitem{henriques2020children}
I.~Henriques and P.~Hartung, ``Children's rights by design in ai development
  for education,'' \emph{The International Review of Information Ethics},
  vol.~29, 2020.

\bibitem{unicef-ai-kids}
\BIBentryALTinterwordspacing
``Executive summary artificial intelligence and children's rights - unicef,''
  2019. [Online]. Available:
  \url{https://www.unicef.org/innovation/reports/memoAIchildrights}
\BIBentrySTDinterwordspacing

\bibitem{lee2020mental}
J.~Lee, ``Mental health effects of school closures during covid-19,'' \emph{The
  Lancet Child \& Adolescent Health}, vol.~4, no.~6, p. 421, 2020.

\bibitem{singh2020impact}
S.~Singh, D.~Roy, K.~Sinha, S.~Parveen, G.~Sharma, and G.~Joshi, ``Impact of
  covid-19 and lockdown on mental health of children and adolescents: A
  narrative review with recommendations,'' \emph{Psychiatry research}, vol.
  293, p. 113429, 2020.

\bibitem{camara2021searching}
A.~C{\^a}mara, N.~Roy, D.~Maxwell, and C.~Hauff, ``Searching to learn with
  instructional scaffolding,'' in \emph{Proceedings of the 2021 Conference on
  Human Information Interaction and Retrieval}, 2021, pp. 209--218.

\end{thebibliography}

\end{document}